# Multidimensional Economic Complexity and Fiscal Crises

Goran Hristovski[1,] Gjorgji Gockov[1] and Viktor Stojkoski[1,2*]

[1]Faculty of Economics, University Ss. Cyril and Methodius, Skopje, North Macedonia

[2]Center for Collective Learning, ANITI, Université de Toulouse & CIAS Corvinus University of Budapest

**Abstract**

Recent studies highlight economic complexity's role in mitigating fiscal crises, often measured via an economy's trade structure. Trade, however, is just one facet of an economy's structure and omits critical innovative activities like research. Here, we investigate how a multidimensional approach to economic complexity—including both trade and research structures—relates to fiscal instability. By using data on over 230 national fiscal crises from 1995 to 2021 and hazard duration analysis, we assess how measures of trade and research complexity combine to explain crisis likelihood. We find that the interaction of complexity dimensions significantly reduces crisis probability, whereas individual indexes alone are not robust predictors. This suggests that economies focusing on a single dimension may be more vulnerable, thus highlighting the importance of balanced development across multiple areas. These findings offer valuable insights for policymakers aiming to enhance economic resilience and mitigate fiscal risks.

[*] Corresponding author vstojkoski@eccf.ukim.edu.mk

1. **Introduction**

In the past two decades, fiscal crises have disrupted the development trajectories of many nations, with significant variation in their severity and frequency (Claessens et al., 2014). These crises underscore the importance of understanding the underlying factors that contribute to a country's vulnerability, particularly within the context of sustainable development. Recent research highlights economic complexity—a measure of the sophistication of a country's productive structure—as a crucial determinant of fiscal stability (El Khattab et al., 2024; Güneri & Yalta, 2021; Hausmann et al., 2014; Hidalgo & Hausmann, 2009; Maggioni et al., 2016; Stojkoski & Kocarev, 2017; Sweet & Eterovic, 2019). Trade-based measures of economic complexity have been found to be inversely correlated with fiscal crises (Gomez-Gonzalez et al., 2023a), i.e., economies exhibiting higher economic complexity—as indicated by sophisticated trade patterns— tend to experience fewer fiscal crises. This suggests that the complexity of a country's export structures could enhance its resilience to external shocks, thereby reducing fiscal vulnerability (Breitenbach et al., 2022; McIntyre et al., 2018).

While trade-based measures of economic complexity have been found to be effective predictors of fiscal crises, the structure of an economy encompasses much more than just trade (Balland et al., 2022). Recent papers have advanced a multidimensional approach to economic complexity, integrating data on trade, patents, and research activities to better explain variations in economic growth, income inequality, and emission intensity (Stojkoski et al., 2023). The relationship between this multidimensional approach and fiscal crises, however, remains an open issue, warranting further exploration.

In this paper, we extend the existing literature by exploring the relationship between a multidimensional approach to economic complexity—focusing specifically on the integration of trade and research dimensions—and fiscal crises. By utilizing a comprehensive dataset covering 131 countries and over 230 fiscal crisis episodes from 1995 to 2021, we employ hazard duration analysis to evaluate how the interaction between these two dimensions of economic complexity relates with the likelihood of fiscal crises. We find that the interaction of trade and research complexity significantly reduces the probability of such events, but individual dimensions of economic complexity are not robust predictors of fiscal crises. We also find that economies focused on a single dimension of complexity may be more vulnerable to fiscal crises, emphasizing the importance of balanced development across multiple dimensions.

But why does the multidimensional approach offer better insights into mitigating fiscal crises?

A fiscal crisis is typically characterized by a variety of indicators, including sovereign debt defaults, large official financing, high inflation, accumulation of domestic arrears, and a loss of market confidence (Manasse et al., 2003; Medas et al., 2018; Reinhart & Rogoff, 2011). These crises result from a range of economic vulnerabilities that go beyond the export structure of an economy. While trade-based measures of economic complexity have been effective in explaining fiscal crises, they have limitations. These measures tend to be biased toward countries that exports are integrated in regional value chains, capturing only one aspect of a country's productive structure, overlooking development in other crucial areas.

Consider the cases of Greece and Belarus. Greece's fiscal crisis in 2010 can be partially attributed to its low ECI (trade), which reflects a less sophisticated export structure. However, this explanation does not hold for Belarus, which, despite a relatively high ECI (trade) rank of #31, still experienced a severe fiscal crisis in 2013. In Belarus's case, its low ECI (research) rank (#100) suggests that the crisis may be linked to weaknesses in other dimensions of its economy. These examples highlight that a country's economic structure extends beyond trade alone, underscoring the importance of studying multidimensional economic complexity when assessing fiscal crises.

These examples emphasizes a comprehensive understanding of fiscal crises requires moving beyond trade-based measures economic complexity (Stojkoski et al., 2023). The productive structure of an economy encompasses not only the export structure of the economy, but also its research capacity, technological innovations, digitalization capabilities, etc.—factors critical for long-term growth, the development of new industries, and resilience to economic shocks (Amarante et al., 2023; Uribe, 2025; Vergara, 2021). Integrating these structures into the economic complexity analysis should provide a deeper understanding of how economies can mitigate fiscal crises. Countries excelling across all dimensions should be better equipped to manage public debt, maintain investor confidence, and reduce vulnerability to fiscal crises, offering a more robust framework for addressing these challenges.

The remainder of this paper is structured as follows: Section 2 reviews the relevant literature, providing a detailed context for our study. Section 3 outlines the data and methodology used in our analysis, detailing our approach to modeling the relationship between economic complexity and fiscal crises. Section 4 presents our findings, whereas Section 5 discusses the implications of our study and suggests directions for future research.

## 2. The Link between Economic Complexity and Fiscal Crises

The literature on fiscal crises offers various definitions and explanations, but there is still no consensus on what precisely constitutes a fiscal crisis (Chakrabarti & Zeaiter, 2014; Detragiache, 2001; Manasse et al., 2003; Medas et al., 2018; Moreno Badia et al., 2022). In theory, fiscal crises emerge when there is a fundamental misalignment between fiscal policy and prevailing macroeconomic conditions, leading to an unsustainable path of public debt accumulation (Cottarelli, 2011). This situation signals a breakdown of fiscal sustainability, typically caused by structural economic imbalances, poor fiscal management, or external shocks.

In practice, however, the definition is more complex. Standard indicators—such as the debt-to-GDP ratio, primary fiscal balance, interest payments as a percentage of revenue, sovereign credit ratings, and bond yield spreads—are commonly used to evaluate a government's fiscal health (Hilscher & Nosbusch, 2010; Petrova et al., 2011). Each of these measures captures a specific aspect of fiscal sustainability, providing insights into a country's capacity to manage its debt burden, meet debt obligations, and maintain economic stability. When these metrics show signs of deterioration, they often signal an increased risk of a fiscal crisis, reflected in higher borrowing costs and restricted access to financial markets (Borio et al., 2023; Moreno Badia et al., 2022).

This traditional approach, however, might be too narrow and fail to capture the full complexity of fiscal crises. For instance, it often overlooks the role of expectations and confidence in financial markets. A government's fiscal position might appear sustainable based on macroeconomic indicators, yet a sudden shift in investor sentiment can trigger a crisis. Moreover, relying solely on these indicators may not be sufficient to predict fiscal crises accurately (Caselli et al., 2022). Some countries have managed high debt levels without experiencing crises, while others have faced crises with relatively moderate debt ratios. This suggests that other factors—such as institutional quality, economic structure, and external vulnerabilities—also play significant roles (Babecký et al., 2014; Frankel & Saravelos, 2012; Giraldo et al., 2023).

Considering these limitations, we follow the approach of (Medas et al., 2018) and define a fiscal crisis as the occurrence of at least one of the following four events: (1) credit events associated with sovereign debt, such as defaults and restructurings; (2) large official financing, like substantial IMF support; (3) implicit domestic public default, through high inflation rates and accumulation of domestic arrears; and (4) loss of market confidence, indicated by loss of market access and high borrowing costs. Together, these criteria form a comprehensive framework for understanding fiscal distress, capturing both the observable symptoms of a fiscal crisis and the underlying structural weaknesses that drive fiscal instability. This definition allows us to consider not only traditional macroeconomic indicators but also broader factors that might affect fiscal stability.

Understanding fiscal crises through this broader perspective opens the door to exploring how structural characteristics of an economy relate with its fiscal health. These structural characteristics are often quantified through measures of economic complexity. These measures estimate the sophistication and value of a country's specialization patterns by applying dimensionality reduction techniques to large datasets (Mealy et al., 2019). Originally developed using international trade data, these metrics assess the complexity of the products that countries export (Hidalgo & Hausmann, 2009). The underlying idea is that countries capable of producing more complex products possess a higher level of collective knowledge and capabilities embedded within their economies. These measures have been validated by their ability to predict future economic growth (Chávez et al., 2017; Hausmann et al., 2014; Koch, 2021; Li & Rigby, 2023; Mighri & AlSaggaf, 2023; Ourens, 2013) and explain geographic variations in inequality (Hartmann et al., 2017; Hartmann & Pinheiro, 2022; Morais et al., 2018; Sbardella et al., 2017);) and emissions (Can & Gozgor, 2017; Neagu, 2019; Peters & Hertwich, 2008; Romero & Gramkow, 2021). Their applicability has been demonstrated across various geographic scales and economic activities, including neighborhoods, cities, regions, and countries, and for activities ranging from urban amenities to patentable technologies (Balland & Rigby, 2017; Boschma et al., 2015; Kogler et al., 2013; Lo Turco & Maggioni, 2022). By capturing the collective knowledge and productive capabilities of an economy, economic complexity provides valuable insights into a country's development potential, structural robustness, and capacity to innovate.

Building on the established role of trade-based measures of economic complexity in explaining growth, inequality, and emissions, recent empirical studies suggest that these measures can also be instrumental in understanding fiscal health. For instance, (Gomez-Gonzalez et al., 2023a) argue that economic complexity is pivotal in maintaining macroeconomic stability by reducing the probability of fiscal crises. Similarly, (El Khattab et al., 2024) emphasize the role of economic complexity in mitigating the adverse impacts of global financial crises. Moreover, (Güneri &

Yalta, 2021) re-assert the negative relationship between economic complexity and output volatility in developing countries, highlighting the importance of diversifying productive structures.

Further evidence of the stabilizing effects is provided by firm-level studies. (Maggioni et al., 2016) demonstrate that firms can reduce output volatility by increasing the complexity of their products. This stabilizing effect is also reflected in the significant positive impact on total factor productivity (Sweet & Eterovic, 2019).

In addition, (Marhubi, 2021) finds that economies with greater complexity tend to experience lower inflation. Financial market indicators support the view for the stabilizing effect of economic complexity as well; (Ã–zmen, 2019) and (Gomez-Gonzalez et al., 2023b) observe that higher economic complexity correlates with lower sovereign yield spreads. (Nguyen, 2022) highlight that higher levels of economic complexity are associated with reductions in the size of the shadow economy. Lastly, (Sarsar & Echaoui, 2024) explore the moderating role of economic complexity in the relationship between energy transition and economic growth, showing that economic complexity positively moderates the negative impact of energy transition on economic growth. In summary, these studies suggest that economic complexity acts as a safeguard against the triggers of various crises by fostering economic resilience and stability.

Despite the growing recognition of economic complexity's importance, most research on its impact on fiscal outcomes has traditionally utilized international trade data (Hausmann et al., 2014). While trade data has been the standard in international comparisons of productive structure, recent advancements suggest adopting a multidimensional approach to economic complexity

(Stojkoski et al., 2023). By combining trade data with patent applications and research publications, (Stojkoski et al., 2023) have developed models that significantly improve the explanatory power of economic complexity metrics. This multidimensional measure explains variations in inclusive green growth, future economic growth, and income inequality, and correlate with lower emission intensities. Most recently, this approach has also been validated by other studies (Baskak, 2024; Cardoso et al., 2023; El Khattab et al., 2024) and extended to even other structural dimensions (Juhász et al., 2024; Stojkoski et al., 2024).

Building on these advancements, we adopt the multidimensional definition of economic complexity to investigate its impact on fiscal crises. By integrating trade data with indicators of innovation and knowledge creation—specifically research publications—we aim to capture a more comprehensive picture of a country's productive capabilities (Stojkoski et al., 2023). This approach aligns with our earlier definition of a fiscal crisis as the multidimensional measure of economic complexity may provide more details about the structural factors that are related with fiscal capacity. This broader perspective allows us not only to investigate how the different dimensions of economic complexity are related with fiscal crisis, but also how the different dimensions interact in explaining the fiscal crisis. In doing so, we contribute to the understanding of fiscal vulnerability beyond traditional macroeconomic indicators, incorporating structural characteristics about innovation that may play a critical role in fiscal stability.

### 3. Data and Methods

*3.1. Methodology*

We model the dependent variable as a binary indicator taking the value of one if a fiscal crisis is observed in each period for a particular country, and zero if not. Then, we employ a hazard duration model, which offers a probabilistic framework for understanding the time until a fiscal crisis occurs, to investigate its relationship with multidimensional economic complexity.

In the context of the Cox PH model, the hazard rate represents the likelihood that a fiscal crisis will occur at a specific time, assuming the country has not yet experienced a crisis. The hazard function is expressed as:

$$\lambda(t; X, ECI, \beta, \lambda_0) = \lambda_0(t) \exp(X, ECI, \beta),$$

where:

$\lambda_0(t)$ represents the baseline hazard function, X is a vector of covariates, including institutional, macroeconomic, and fiscal variables. As specific variables we include the Economic Complexity Index's (ECIs) measured through trade and research data, whereas $\beta$ denotes the vector of parameters to be estimated.

For this study, we employ the semi-parametric Cox Proportional Hazards model (Cox, 1972). Under this specification, the coefficients in the Cox model represent the constant, proportional impact of each covariate on the likelihood that a country will enter a fiscal crisis at a given point in time. Covariates are set before the occurrence of the event, meaning they are determined prior to the onset of the fiscal crisis. The timing of the event reflects the point when these covariates are fixed.

In some cases, ties in events may occur, meaning multiple economies experience a fiscal crisis simultaneously. This presents a challenge because the standard Cox proportional hazards model assumes unique event times. When ties arise, they complicate the estimation of model parameters and hazard ratios. To address this, we additionally perform robustness check by estimating the model parameters using the Efron and Breslow methods. (Gomez-Gonzalez et al., 2023a).

### 3.2. Data

This research draws on an extensive dataset covering 131 countries and over 230 fiscal crisis episodes from 1998 to 2021. Key variables include the Economic Complexity Index (ECI) for Trade and Research, sourced from the Observatory of Economic Complexity (OEC, Simoes and Hidalgo, 2011), Regulatory Quality and Rule of Law indicators from the World Bank database, and data on Real GDP Growth and Interest Expenses as a percentage of GDP from the World Economic Outlook (WEO) (see Table 1 for data source).

The trade data used to calculate the ECI are based on international trade records, which classify goods into categories according to the Harmonized System at the 4-digit level (HS4), whereas the research publication data come from the SCImago Journal & Country Rank, which tracks country-level statistics on documents published in Scopus. Moreover, the standard estimation of multidimensional economic complexity also includes the technology dimension estimated using patent data. This data, however, is limited in availability for many to less than 100 countries (the trade and research dimension contain data for more than 130 countries). Due to this, we exclude it from our models.

The descriptive statistics of key variables are shown in Table 1. They exhibit a high variability across the sample. The mean real yearly GDP growth rate is 3.58%, with a wide range from -66.66% to 177.26%, highlighting divergent economic performance among countries. Interest expenses as a percentage of GDP average 1.46%, though the range from -35.48% to 17.71% indicates differing debt burdens. Both regulatory quality and rule of law have negative mean values (-0.18 and -0.19, respectively), reflecting weaker governance in many countries, with notable variation as reflected in their standard deviations. The Economic Complexity Indices for Trade and Research show moderate levels of complexity (mean values of 0.53 and 0.54, respectively), with a moderate degree of variation across the sample. These statistics underscore the diverse economic environments within the dataset, which is crucial for understanding the relationship between economic complexity and fiscal crises.

**Table 1. Summary Statistics.**

| Indicator | Abbreviation | Source | Median | Mean | Std.Dev | Max. | Min. |
|---|---|---|---|---|---|---|---|
| Real GDP growth | growth | WEO | 3.32 | 3.58 | 6.39 | 177.26 | -66.66 |
| Interest expenses as % GDP | interest | WEO (estimate) | 1.9 | 1.46 | 2.45 | 17.71 | -35.48 |
| Regulatory quality | rqe | World Bank | -0.09 | -0.18 | 1.19 | 6.47 | -5.78 |
| Rule of law | rle | World Bank | -0.06 | -0.19 | 1.13 | 5.84 | -4.41 |
| Economic Complexity Index Trade | eciT | OEC | 0.55 | 0.53 | 0.21 | 1 | 0.02 |
| Economic Complexity Index Research | eciR | OEC | 0.56 | 0.54 | 0.21 | 1 | 0.09 |

## 4. Results

### 4.1. Main findings

We analyze the relationship between multidimensional ECI and fiscal crises by constructing ten different models. Each of these models examines various combinations of trade and research complexities, both individually and through interactions, while controlling for other relevant economic and institutional factors. Table 2 provides our regression analyses results.

In the initial set of models (Models 1 to 5), we assess the effects of ECI (trade) and ECI (research) without control variables. Both trade and research complexities individually show a significant negative effect on the probability of fiscal crises. When combined, however, their individual effects become statistically insignificant and reverse direction, while their interaction shows a significant inverse relationship with fiscal crises. These models demonstrate have a limited explanatory power, as reflected by their AIC values, highlighting the importance of including potential economic and institutional factors.

We shift our focus to selecting the most appropriate model by progressively incorporating control variables and assessing model fit using the Akaike Information Criterion (AIC), where lower AIC values indicate better models (Farooq & Karami, 2019). We begin with the baseline model (Model 6), which includes only the control variables: regulatory quality, interest expenses as a percentage of GDP, real GDP growth, and rule of law. The AIC for this model is 708.20. Among the controls, interest expenses and rule of law appear significant predictors of fiscal crises. Higher interest expenses increase the probability of a fiscal crisis, while stronger rule of law reduces it. Regulatory quality and real GDP growth do not exhibit statistically significant effects.

In Model 7, we add ECI (trade) to the baseline model. This inclusion results in an improved fit, with the AIC decreasing to 704.06. The negative effect of trade complexity on fiscal crisis probability is significant, suggesting that higher trade complexity reduces fiscal crisis risk even after controlling for other factors, as was also found in (Gomez-Gonzalez et al., 2023a).

Next, we incorporate ECI (research) with the control variables in Model 8. We find that this model does not improve the fit compared to the baseline, as the AIC increases to 710.03. Additionally, research complexity alone appears to not be a significant predictor of fiscal crises when controlling for other variables.

We then explored the combined effect of trade and research complexities in Model 9 by including their interaction term along with the control variables. This model results in a better fit than the baseline, with the AIC decreasing to 703.46. The interaction between ECI (trade) and ECI (research) is significant, demonstrating that the synergy between trade and research complexities plays a crucial role in reducing fiscal crisis risk within a controlled framework.

Finally, in Model 10, we include both ECI variables, their interaction term, and the control variables. This model yields the lowest AIC value of 699.17, indicating the best fit among all models. The results show that research complexity alone may increase fiscal crisis risk, but the combination of trade and research complexities substantially reduces that risk. The significant negative interaction between ECI (trade) and ECI (research) underscores the importance of a multidimensional approach to economic complexity (Stojkoski et al., 2023).

## Table 2. Regression Results.

| Predictors | 1 Effect | 2 Effect | 3 Effect | 4 Effect | 5 Effect | 6 (Baseline) Effect | 7 (ECI(trade)) Effect | 8 (ECI(research)) Effect | 9 (ECI(Multidimensional)) Effect | 10 (Final) Effect |
|---|---|---|---|---|---|---|---|---|---|---|
| ECI (trade) | -3.28 *** (-4.32 – -2.23) | | -2.94 *** (-4.25 – -1.64) | 2.62 (-0.92 – 6.16) | | | -1.87 * (-3.38 – -0.37) | | | 2.78 (-0.86 – 6.41) |
| ECI (research) | | -2.42 *** (-3.43 – -1.40) | -0.56 (-1.87 – 0.74) | 4.14 ** (0.99 – 7.28) | | | | -0.28 (-1.59 – 1.04) | | 4.38 ** (1.16 – 7.60) |
| ECI (trade) x ECI (research) | | | | -10.04 ** (-16.17 – -3.91) | -3.73 *** (-4.94 – -2.52) | | | | -2.33 * (-4.14 – -0.51) | -9.27 ** (-15.77 – -2.77) |
| Regulatory quality | | | | | | -0.28 (-0.80 – 0.24) | -0.07 (-0.63 – 0.48) | -0.28 (-0.80 – 0.24) | -0.17 (-0.70 – 0.36) | -0.10 (-0.66 – 0.46) |
| Interest expenses as % GDP | | | | | | 0.15 *** (0.07 – 0.23) | 0.16 *** (0.08 – 0.24) | 0.15 *** (0.07 – 0.23) | 0.16 *** (0.08 – 0.25) | 0.16 *** (0.08 – 0.25) |
| Real GDP growth | | | | | | -0.02 (-0.09 – 0.05) | -0.03 (-0.10 – 0.04) | -0.02 (-0.09 – 0.05) | -0.03 (-0.10 – 0.04) | -0.04 (-0.11 – 0.03) |
| Rule of law | | | | | | -0.84 ** (-1.35 – -0.32) | -0.74 ** (-1.26 – -0.21) | -0.80 ** (-1.35 – -0.25) | -0.58 * (-1.13 – -0.02) | -0.62 * (-1.19 – -0.05) |
| Observations | 221 | 221 | 221 | 221 | 221 | 221 | 221 | 221 | 221 | 221 |
| AIC | 728.636 | 748.019 | 729.918 | 719.946 | 723.732 | 708.203 | 704.061 | 710.034 | 703.464 | 699.174 |
| log-Likelihood | -363.318 | -373.010 | -362.959 | -356.973 | -360.866 | -350.102 | -347.031 | -350.017 | -346.732 | -342.587 |

* p<0.05   ** p<0.01   *** p<0.001

In Figure 1 we summarize both the selection of the final model and the marginal effects of ECI (trade), ECI (research), and their interaction on the probability of fiscal crises. The negative and significant interaction term indicates that countries with high levels of both trade and research complexities have a substantially lower probability of experiencing fiscal crises. Conversely, research complexity alone is positively associated with the probability of fiscal crises. This suggests that while research complexity by itself may increase vulnerability, the combination with trade complexity mitigates this risk and strengthens fiscal stability.

The analysis reveals that while trade complexity alone helps reduce the risk of fiscal crises, its effect is not consistently robust across all models. The mitigating impact of trade complexity, however, is significantly amplified when combined with research complexity, underscoring the critical role of innovation in fostering resilience. Although research complexity on its own does not reliably predict fiscal crises and may even indicate increased vulnerability in certain contexts, its true value becomes evident when integrated with trade sophistication. Therefore, innovation is central to enhancing a country's resilience, enabling it to increase scientific output and export more sophisticated products—both of which contribute to fiscal stability (Anthopoulos & Siozos, 2015; Fukao et al., 2009; Grossman & Helpman, 1989; Tellis et al., 2008; Uribe, 2025). Together, they reflect a country's structural transformation and innovation capacity, the key drivers of long-term resilience.

The control variables offer additional insights besides the ones provided by the economic complexity measures. Interest expenses consistently show a positive and highly significant effect across all models including controls, highlighting the critical role of sound fiscal management in preventing fiscal crises. The significance of the rule of law demonstrates that strong legal institutions and enforcement mechanisms are essential in reducing fiscal crisis risk. Regulatory quality and real GDP growth did not have significant effects within this analysis.

**Figure 1. Evaluating Model Fit Using AIC for Multidimensional Economic Complexity**

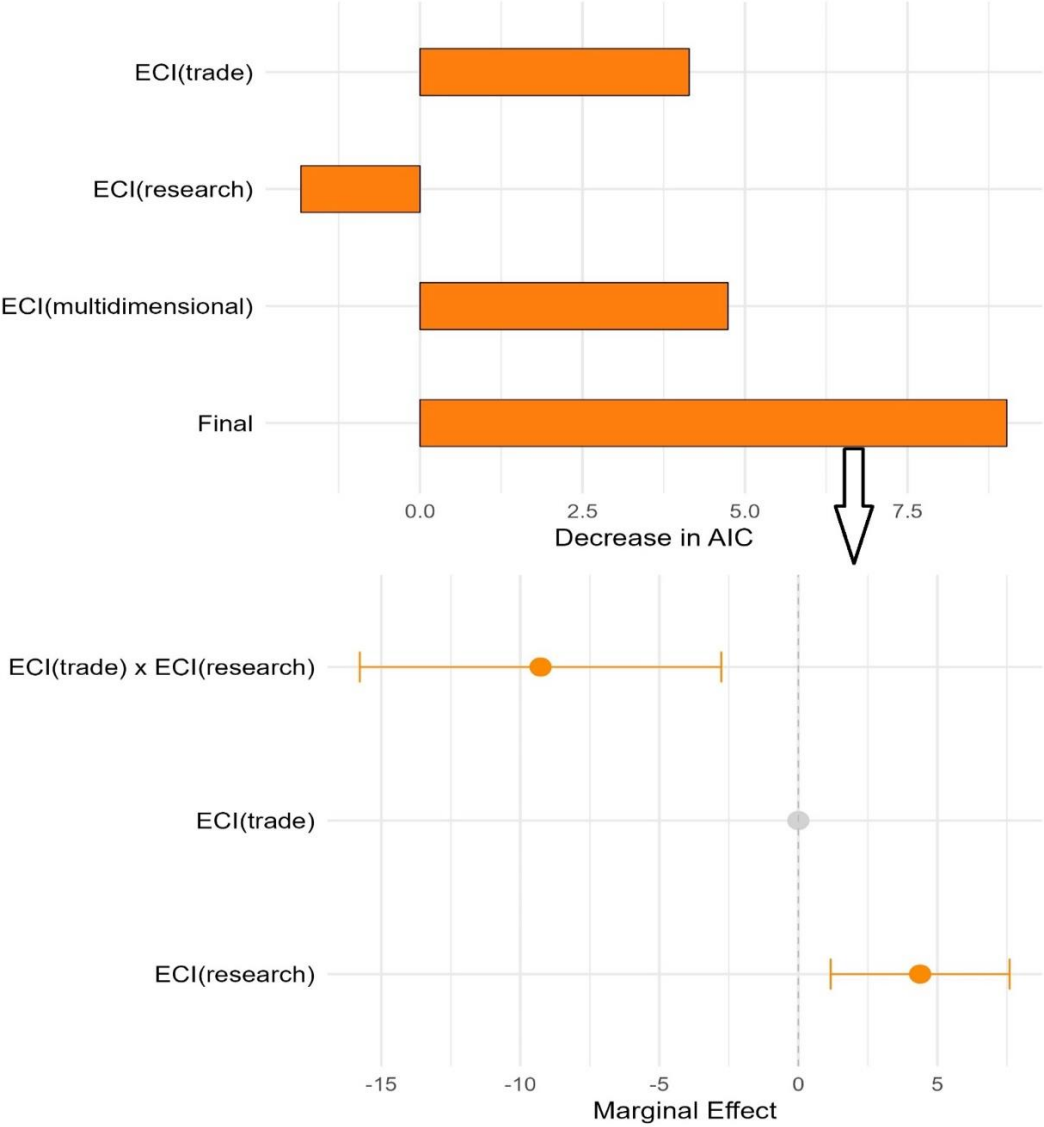

### 4.2. Robustness check

To further ensure the robustness of our findings, we conduct additional checks using Efron's and Breslow's approximations to handle tied event times in the survival data. The Efron approximation is usually preferred when the sample size is small either from the outset or due to heavy censoring, while the Breslow approximation is particularly effective in scenarios with a moderate or heavy ties (Hertz-Picciotto & Rockhill, 1997). We apply both methods to ensure a comprehensive analysis. In Table 3 we display the results.

These results indicate that our main findings remain consistent across these robustness checks, reinforcing the validity and reliability of our conclusions. Specifically, we observe a marginal reduction in the estimated effects of the interaction between economic complexity dimensions and the individual economic complexity measures in both the Breslow model (Column 2 of Table 3) and the Efron Model (Column 3 of Table 3). There are slight differences, however, in the marginal effects of the control variables. Namely, interest expenses slightly decrease but remain statistically significant, while Rule of Law becomes statistically insignificant. All other covariates remain the same. These results underscore the robustness of our model, affirming that the core relationships identified are not sensitive to the choice of approximation methods used for handling tied event times.

**Table 3. Regression Robustness Checks.**

| Predictors | Final model | Breslow | Efron |
|---|---|---|---|
| ECI (research) | 4.38 ** (1.16 – 7.60) | 3.56 * (0.85 – 6.28) | 3.58 ** (0.93 – 6.23) |
| ECI (trade) | 2.78 (-0.86 – 6.41) | 2.54 (-0.68 – 5.77) | 2.55 (-0.65 – 5.76) |
| ECI (trade) x ECI (research) | -9.27 ** (-15.77 – -2.77) | -7.83 ** (-13.48 – -2.17) | -8.08 ** (-13.68 – -2.48) |
| Real GDP growth | -0.04 (-0.11 – 0.03) | -0.03 (-0.09 – 0.03) | -0.03 (-0.09 – 0.03) |
| Regulatory quality | -0.10 (-0.66 – 0.46) | -0.11 (-0.61 – 0.39) | -0.16 (-0.66 – 0.34) |
| Interest expenses as % GDP | 0.16 *** (0.08 – 0.25) | 0.12 *** (0.06 – 0.18) | 0.14 *** (0.07 – 0.20) |
| Rule of law | -0.62 * (-1.19 – -0.05) | -0.51 (-1.03 – 0.00) | -0.50 (-1.02 – 0.01) |
| Observations | 221 | 221 | 221 |
| AIC | 699.174 | 1317.208 | 1285.065 |
| log-Likelihood | -342.587 | -651.604 | -635.532 |

## 5. Discussion

In this paper, we investigated the role of multidimensional economic complexity measured through the trade and research dimensions in mitigating fiscal crises. We found that, while individual dimensions of trade and research alone do not serve as robust predictors of fiscal crises, their interaction significantly explains the probability of a country to mitigate a fiscal crisis. This result suggests that diversified economies, which focus on a variety of specializations, rather than just one dimension, could be vital for achieving structural resilience and fiscal stability.

By advancing our understanding of how economic complexity relates to fiscal stability, this study contributes substantially to the broader literature on financial crises and economic resilience. Policymakers can leverage these insights to design more effective strategies for mitigating fiscal risks and enhancing economic robustness. Strategies may include fostering balanced development of multiple dimensions of the economic structure, thus sustaining real GDP growth, keeping interest expenses under control to preserve fiscal space, and strengthening institutional frameworks such as regulatory quality, rule of law.

Nevertheless, despite the robustness of our findings, the study has certain limitations. For once, the inclusion of the technology dimension was limited by a reduction in available observations, which may have led to an underestimation of its impact (Ivanova et al., 2017; Juhász et al., 2024; Lybbert & Xu, 2022; Stojkoski et al., 2016, 2024). Similarly, data on the digital economy covers a shorter time span, preventing us from fully capturing this structural dimension. Future research should aim to incorporate a broader set of dimensions, which would better leverage the power of economic complexity in explaining fiscal crises. We anticipate that this will become feasible as more comprehensive data becomes available in the near future.

Similarly, our study focused exclusively on fiscal crises and did not account for other types of financial crises, such as banking and currency crises, which may also affect fiscal stability (M. D. Bordo & Meissner, 2006, 2016; M. Bordo & Eichengreen, 1999; Reinhart & Rogoff, 2009). Extending the analysis to include these crises would offer a more comprehensive understanding of fiscal vulnerabilities and the interplay between different types of economic shocks.

Despite these limitations, our study has important implications to the impact of the economic structure on the ability of a nation to mitigate a financial crisis. It calls for a balanced approach to economic development and provides a foundation for future research to further elucidate the relationship between economic complexity and financial stability.

**Disclosure statements**

The authors report there are no competing interests to declare.

**Acknowledgments**